\documentclass[showpacs,prl,twocolumn]{revtex4}

\usepackage{epsfig,amsmath}
\usepackage{subfigure}
\usepackage{graphicx}
\usepackage{dcolumn}
\usepackage{stmaryrd}
\usepackage{mathrsfs}
\usepackage{pifont}
\usepackage{amsthm}
\usepackage{amssymb}
\usepackage{bm}
\usepackage{latexsym}
\usepackage{hyperref}
\usepackage{color}
\usepackage{appendix}
\usepackage{braket}
\usepackage{amsmath}
\usepackage{array}
\usepackage{algorithm}
\usepackage{algpseudocode}

\DeclareUnicodeCharacter{0301}{\'{e}}

\newcommand{\beq}{\begin{equation}}
	\newcommand{\eeq}{\end{equation}}
\newcommand{\beqa}{\begin{eqnarray}}
	\newcommand{\eeqa}{\end{eqnarray}}

\begin{document}
	
	\title{End-to-End Learning of Quantum Control on Latent Dynamical Manifold}
	
	\begin{abstract}
		Traditional quantum control relies on an iterative “simulate-then-optimize” paradigm, where dynamics simulation and control design are decoupled, leading to substantial computational overhead and limited scalability, particularly in noisy environments.
		Here, we propose an end-to-end quantum control framework based on long
		short-term memory, in which system dynamics and control strategies are learned jointly in a low
		dimensional latent manifold. The model directly maps initial states and environmental parameters
		to both dynamical trajectories and optimized control pulse in a single forward pass. 
		The framework is validated on adiabatic speedup in a two-level system and state transfer in a one-dimensional spin chain under noise, achieving accurate dynamical prediction and control optimization. It improves the fidelity for both tasks and significantly reduces the optimization cost by three orders of magnitude compared with conventional iterative methods, while exhibiting strong generalization to multi-parameter, time-varying noise, as well as to different initial states and driving fields.
		Our work introduces a data-driven control paradigm based on latent manifold learning, reducing the computational bottleneck of iterative optimization and enabling real-time adaptive control of complex open quantum systems.
	\end{abstract}
	
	\author{Jun-Dong Zhong$^{1}$, Zong-Yuan Ge$^{1}$,Feng-Hua Ren$^{2}$\footnote{renfenghua@qtu.edu.cn}, Zhao-Ming Wang\textsuperscript{1,3,4}\footnote{wangzhaoming@ouc.edu.cn}}
	\address{$^{1}$College of Physics and Optoelectronic Engineering, Ocean University of China, Qingdao 266100, China}
	\address{$^{2}$School of Information Management and School of Artificial Intelligence, Qingdao University of Technology, Qingdao 266520, China}
	\address{$^{3}$Engineering Research Center of Advanced Marine Physical Instruments and Equipment of Ministry of Education, Qingdao 266100, China}
	\address{$^{4}$Qingdao Key Laboratory of Optics and Optoelectronics, Qingdao 266100, China}
	
	\date{\today}
	\maketitle
	
	\textit{Introduction---}Precise quantum control is a cornerstone of quantum computing~\cite{Bluvstein2023}, quantum metrology~\cite{Giovannetti2006}, and quantum simulation~\cite{Altman2021}. 
	In practice, environmental noise and intrinsic control limitations degrade performance, making the design of stable control strategies a central challenge~\cite{Poggi2024,Ding2025,gao2026supervised,Shi2024}, particularly under time-varying noise. 
	Among various control strategies~\cite{ZHANG20231,PhysRevApplied.23.054002,Shi2024}, the leakage elimination operator (LEO) approach is widely used to suppress transitions outside the target subspace~\cite{PhysRevLett.114.190502}. The corresponding LEO Hamiltonian can be realized by a sequence of zero-area pulses~\cite{Wang201803}, with exact control pulse conditions derived via the Feshbach PQ partitioning technique in closed systems~\cite{Wang2020a}.
	However, its performance degrades with increasing system-bath coupling~\cite{Wang2020b}, requiring additional pulse refinement strategies~\cite{Xie2022}.
	Nevertheless, traditional optimized control methods such as Gradient Ascent Pulse Engineering~\cite{Khaneja2005}, Chopped Random Basis method~\cite{Caneva2011}, and Krotov~\cite{Floether2012,Fernandes2023}, as well as gradient-based optimization schemes (e.g., Adam~\cite{Xie2022}) combined with backpropagation through time (BPTT)~\cite{werbos1990backpropagation}, typically rely on the ``simulate-then-optimize'' paradigm. This leads to rapidly increasing computational cost with system size and limits scalability, as repeated numerical integration of the Lindblad master equation and BPTT are required, while failing to exploit the temporal structure of quantum evolution for adaptive control.

	Recently, neural network-based machine learning methods have been widely applied to simulate open quantum system dynamics, including physics-informed neural networks (PINNs)~\cite{LiZhenyu2025}, variation methods~\cite{bond2024open, guo2025variational}, and recurrent neural networks (RNNs)~\cite{Banchi2018,https://doi.org/10.48550/arxiv.2401.06380,LuoDi2022}. RNNs, particularly long short-term memory (LSTM)~\cite{Mienye2024,an2025dual}, have been used to learn low-dimensional latent representations of quantum dynamics~\cite{lusch2018deep, rubanova2019latent, krishnan2015deep}, achieving success in modeling open many-body systems~\cite{Lin2021}. These approaches offer reduced computational cost compared with direct numerical integration and enable extrapolation beyond the training time window~\cite{Mohseni2022}.
	
	However, in the RNN-based approaches, control is still treated separately via iterative optimization~\cite{zhong2026optimal}, thereby inheriting the same ``simulate-then-optimize'' paradigm and failing to fully leverage the representational power of RNNs to unify dynamics and control. 
	In contrast, PINNs approaches provide a unified framework that couples dynamics and control within a single optimization objective~\cite{Norambuena2024}, but their applicability is typically limited to interpolation within the training distribution, with poor generalization to unseen, time-varying, or multi-parameter variations, necessitating retraining under new operating conditions.
	
	\begin{figure}[htbp]
		\centering
		\includegraphics[width=\linewidth]{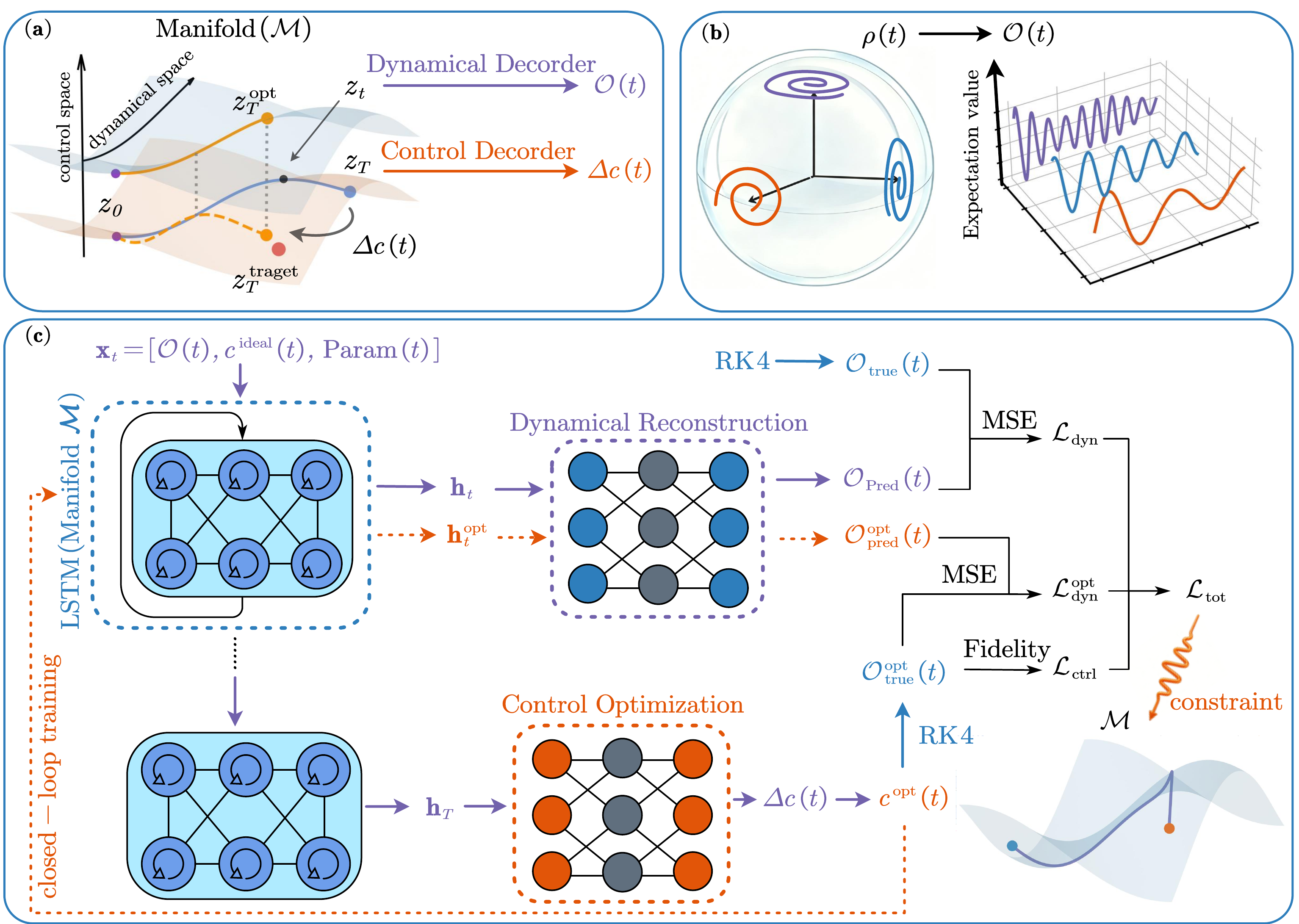}
		\caption{End-to-end quantum control framework based on latent dynamical manifold learning. 
			(a) Schematic illustration of the proposed latent manifold $\mathcal{M}$. Different control pulses give rise to distinct effective dynamical regimes, which are visualized as separate surfaces for clarity. 
			(b) Projection of the quantum states onto local observables.
			(c) End-to-end training framework. The LSTM hidden states sequence $\mathbf{h}_t$ forms a latent trajectory $z_t$ in $\mathcal{M}$. A dynamical decoder maps $\mathbf{h}_t$ to observables, and a control decoder maps the final state $\mathbf{h}_T$
			to a control correction added to the ideal control pulse. The representation $\mathcal{M}$ is learned jointly through dynamical prediction loss $\mathcal{L}_{\mathrm{dyn}}$, control optimization loss $\mathcal{L}_{\mathrm{ctrl}}$, and closed-loop dynamical consistency loss $\mathcal{L}^{\mathrm{opt}}_{\mathrm{dyn}}$.}
		\label{Fig.1}
	\end{figure}
	
	To address this limitation, we propose an end-to-end quantum control framework based on latent dynamical manifold learning, which constructs a control-conditioned latent representation that jointly encodes dynamics and control, enabling strong generalization across multi-parameter, time-varying environments and varying initial states and driving fields without retraining.
	We validate the framework on two examples in noisy environments, including adiabatic speedup in a two-level system and quantum state transfer in a one-dimensional spin chain.
	The results show that the proposed framework enables accurate dynamical prediction and efficient control optimization while significantly improving computational efficiency compared with conventional iterative approaches.
	
	\textit{Joint Learning of Quantum Dynamics and Control via Latent Manifold---}Figure~\ref{Fig.1} presents the end-to-end framework for quantum dynamical prediction and control optimization based on latent dynamical manifold learning, where an LSTM-based generator embeds both dynamics and control into a unified latent manifold $\mathcal{M}$~\cite{Lin2021,Banchi2018,Norambuena2024} equipped with dual decoders. 
	
	Here, $\mathcal{M}$ provides a compact latent representation of quantum trajectories, in which states with similar dynamics are mapped to nearby regions.
	Figure~\ref{Fig.1}(a) provides a schematic illustration of the latent manifold. Different control pulses give rise to distinct dynamical regimes, which are visualized as separate surfaces for clarity. Within each regime, quantum trajectories evolve in the latent dynamical space, while the latent state $z_t$ encodes both dynamical evolution and control information. Optimized control drives the system toward trajectories with higher fidelity.
	
	Open quantum systems are commonly described by the Lindblad master equation~\cite{Manzano2020}, while non-Markovian dynamics~\cite{Yu2004,Wang2021jpa,Strunz1999} capture memory effects arising from structured bosonic environments. For detail, see Supplemental Material (SM) S1. To avoid the exponential cost of propagating the full density matrix, the quantum state is projected onto local observables (Fig.~\ref{Fig.1}(b)), yielding the expectation values $\mathcal{O}(t)=\langle \mathcal{O}(t)\rangle$. Local observables suffice, as the control objective (e.g., population transfer) depends only on local expectations, and the LSTM implicitly captures non-local correlations in its hidden and cell states~\cite{elman1990finding,hochreiter1997long,greff2016lstm}.
	We further construct the input sequence $\mathbf{x}_t = [\mathcal{O}(t),\; c(t),\; \mathrm{Param}(t)]$, where the initial observable $\mathcal{O}(0)$ is taken from the true trajectory and $\mathcal{O}(t>0)$ is generated autoregressively as $\mathcal{O}_{\mathrm{pred}}(t)$, $c(t)$ denotes the control pulse.
	For $\mathrm{Param}(t)$, unlike conventional approaches assuming static or slowly varying noise~\cite{Kestner2013,Rong2014}, we treat it as time-varying inputs, enabling the LSTM to learn a mapping from instantaneous environmental conditions to quantum dynamics and optimized control, consistent with realistic noise such as NV electron spins coupled to $^{13}C$ nuclear spin baths~\cite{Xiedu2023}.
	
	In contrast to conventional latent representations designed solely for dynamical prediction~\cite{zhong2026optimal}, the LSTM induces a latent representation via its hidden states, which jointly encodes system dynamics and control information. As illustrated in Fig.~\ref{Fig.1}(c), the hidden-state sequence $\mathbf{h}_t$ forms a trajectory $z_t$ in $\mathcal{M}$, providing an internal representation of the quantum dynamics.
	A dynamical decoder $\mathcal{D}_{\mathrm{dyn}}$ maps $\mathbf{h}_t$ to observable expectation values  $\mathcal{O}_{\mathrm{pred}}(t+1)=\mathcal{D}_{\mathrm{dyn}}(\mathbf{h}_{t+1})$~\cite{hornik1989multilayer},
	where $\mathbf{h}_{t+1}$ is generated autoregressively via
	$(\mathbf{h}_{t+1},\mathbf{c}_{t+1})=\mathrm{LSTM}(\mathbf{x}_{t+1},\mathbf{h}_t,\mathbf{c}_t)$~\cite{https://doi.org/10.48550/arxiv.1909.09586, Krichen2025}. For detail, see SM S3.
	Together, the LSTM evolution and dynamical decoder define a learned latent-space dynamics that predicts the observable trajectories of the quantum system.
	For control optimization, a control decoder $\mathcal{D}_{\mathrm{ctrl}}$ maps the final hidden state $\mathbf{h}_T$, which encodes the global system dynamical evolution, to a control coefficient vector $\mathbf{I} = \mathcal{D}_{\mathrm{ctrl}}(\mathbf{h}_T)$ that parameterizes the optimized control pulse,
	\begin{equation}
		c^{\mathrm{opt}}(t)=c^{\mathrm{ideal}}(t)+\sum_{k=N}^{M}\mathbf{I}_k\sin[(k+1)\omega t],
		\label{ct}
	\end{equation}
	where $\mathbf{I}_k$ are coefficients of the sinusoidal correction terms added to the ideal control pulse $c^{\mathrm{ideal}}(t)$, enabling flexible frequency-domain shaping. The ideal control pulse is realized via a sequence of zero-area pulses, with the corresponding pulse conditions derived theoretically using the P–Q partitioning technique for closed systems~\cite{Wang2020a}.
	
	As shown in Fig.~\ref{Fig.1}(c), the framework jointly learns dynamics prediction and control generation, together with closed-loop dynamical consistency in the latent representation. The total training objective is given by
	\begin{equation}
		\mathcal{L}_{\mathrm{tot}}=\mathcal{L}_{\mathrm{dyn}}+\mathcal{L}_{\mathrm{ctrl}}+\mathcal{L}_{\mathrm{dyn}}^{\mathrm{opt}}.
	\end{equation}
	The dynamics prediction loss $\mathcal{L}_{\mathrm{dyn}}$ is defined as the mean squared error (MSE)
	\begin{equation}
		\mathcal{L}_{\mathrm{dyn}} =
		\frac{1}{T}
		\sum_{t=1}^{T}
		\left\|
		\mathcal{O}_{\mathrm{pred}}(t)
		-
		\mathcal{O}_{\mathrm{true}}(t)
		\right\|_2^2,
	\end{equation}
	where $\mathcal{O}_{\mathrm{true}}(t)$ is obtained via numerical integration. This loss enforces accurate prediction of quantum dynamics, thereby encoding its dependence on control and environmental parameters in the latent representation. 
	
	The control optimization loss is defined via a fidelity objective as
	\begin{equation}
		\mathcal{L}_{\mathrm{ctrl}} = 1 - F(c^{\mathrm{opt}}(t))+\mathcal{P}_{\mathrm{bound}}(c^{\mathrm{opt}}(t)),
	\end{equation}
	where $F$ denotes the fidelity obtained  from differentiable numerical integration~\cite{chen2018neural,rackauckas2020universal}. This objective trains $\mathcal{D}_{\mathrm{ctrl}}$ to maximize the fidelity with respect to the target state, while $\mathcal{P}_{\mathrm{bound}}$ enforces $|c(t)| \le c_{\max}$ to ensure physical implementability~\cite{Xie2022}.
	
	Feeding back $c^{\mathrm{opt}}(t)$ into the model generates a closed-loop trajectory $\mathbf{h}_t^{\mathrm{opt}}$, with $\mathcal{O}_{\mathrm{pred}}^{\mathrm{opt}}(t)=\mathcal{D}_{\mathrm{dyn}}(\mathbf{h}_t^{\mathrm{opt}})$. The corresponding closed-loop dynamical consistency loss is
	\begin{equation}
		\mathcal{L}^{\mathrm{opt}}_{\mathrm{dyn}} =
		\frac{1}{T}
		\sum_{t=1}^{T}
		\left\|
		\mathcal{O}^{\mathrm{opt}}_{\mathrm{pred}}(t)
		-
		\mathcal{O}^{\mathrm{opt}}_{\mathrm{true}}(t)
		\right\|_2^2.
	\end{equation}
	This loss enforces closed-loop consistency of the latent dynamics under optimized control, improving generalization to distribution shifts induced by optimized control~\cite{quinonero2008dataset}.
	
	\textit{Model Validation---}We first consider adiabatic speedup in an open two-level quantum system. The system is described by the Hamiltonian $H_s(t) = [1-s(t)]H_i + s(t)H_f$, where $s(t)$ is a driving field that increases linearly from 0 to 1~\cite{Farhi2001,Lin2020}. Here we choose $H_i=\sigma_z$, $H_f=\sigma_x$, and set the Lindblad operator to $L=\sigma_-$ for dissipation.
	With LEO control~\cite{Wu2002}, the system Hamiltonian is $H_c(t) = [1+c(t)][(1-s(t))\sigma_z + s(t)\sigma_x]$.  For the sinusoidal ideal control pulse $c^{\mathrm{ideal}}(t)=I\sin(\pi t /\tau)$ where amplitude $I \approx 54.4$, half-period $\tau = 1/2$, the ideal pulse condition is satisfied, corresponding to the third root of the Bessel function $J_0(I\tau/\pi)$~\cite{Wang2020a}. For details, see SM~S2. 
	The adiabatic fidelity is defined as $F(t)=\sqrt{\langle E_0(t)|\rho(t)|E_0(t)\rangle}$, where $|E_0(t)\rangle$ is the instantaneous ground state of $H_s(t)$ and $\rho(t)$ is obtained from local observables~\cite{nielsen2010quantum}.

	\begin{figure}[htbp]
		\centering
		\includegraphics[width=\linewidth]{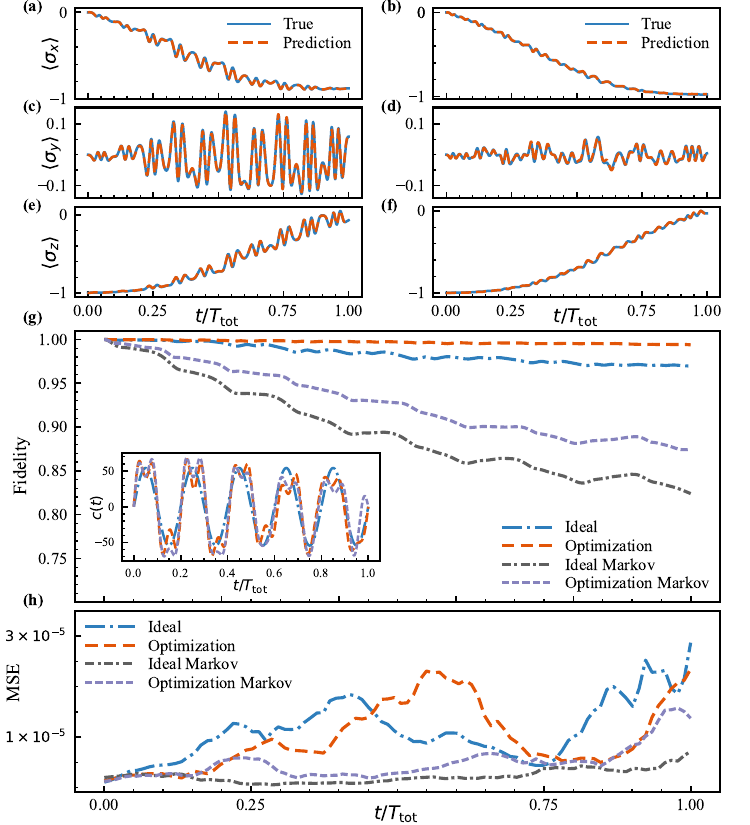}
		\caption{	
			Adiabatic speedup in a two-level open quantum system ($\Gamma=0.03$, $\gamma=1.5$, $T=10$, $T_{\mathrm{tot}}=5$).  
			(a)–(f) Local observables $\langle \sigma_{x,y,z} \rangle$ comparing numerical solutions with model predictions. Left column: ideal control; right column: optimized control. 
			(g) Adiabatic fidelity under ideal and optimized control for non-Markovian ($\gamma=1.5$) and near Markovian ($\gamma=20$) cases. The inset shows the corresponding control pulses. 
			(h) MSE averaged over local observables.}
		\label{Fig.2}
	\end{figure}
	
	\begin{figure}[htbp]	
		\centering
		\includegraphics[width=\linewidth]{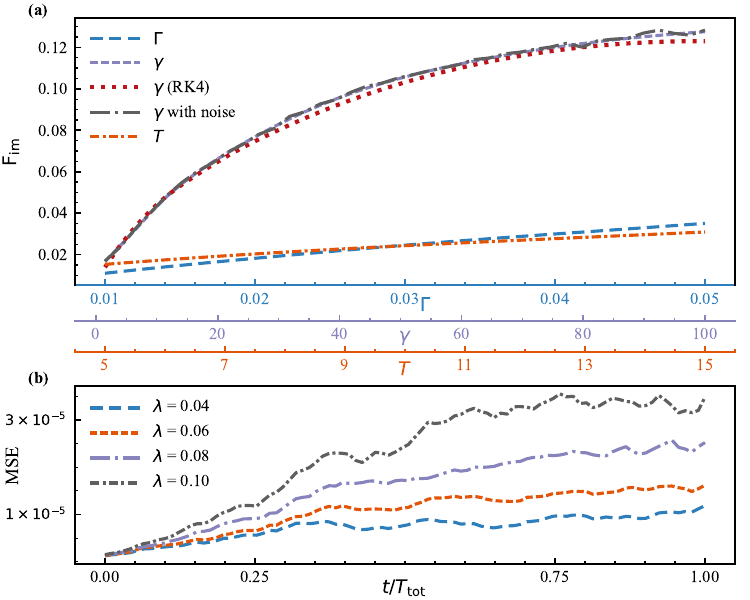}
		\caption{Parameter generalization and stability of the proposed framework under time-varying noise. All results are obtained with the baseline parameters $\Gamma=0.03$, $\gamma=4$, $T=10$, and $T_{\mathrm{tot}}=5$, unless otherwise stated. 
			(a) Final-state fidelity improvement $F_{\mathrm{im}}$ as a function of $\Gamma$, $T$, and $\gamma$, where only the scanned parameter is varied. For the $\gamma$ scan, RK4 benchmarks and results with noise amplitude $\lambda=0.05$ are also included. (b) MSE averaged over environmental parameters under stochastic noise with different amplitudes.}
		\label{Fig.3}
	\end{figure}
	
	Figure~\ref{Fig.2} validates the proposed framework for adiabatic speedup in a two-level open quantum system. Here $\Gamma$, $\gamma$ and $T$ denote the system-bath coupling strength, the bath character frequency, and temperature, respectively~\cite{Wang2021jpa}, for details, see SM~S1. Figure~\ref{Fig.2}(a)–(f) compare the time evolution of $\mathcal{O}(t)$ ($\mathcal{O}=\left\langle \sigma_{x,(y,z)} \right\rangle$) under ideal (left column) and optimized (right column) control, $\Gamma=0.03$, $\gamma=1.5$, $T=10$, and total evolution time $T_{\mathrm{tot}}=5$.
	The predicted dynamics agree closely with fourth-order Runge–Kutta (RK4)~\cite{Yan2017} reference solutions over the entire time interval, demonstrating that the latent trajectories provide a faithful representation of the open system dynamics.
	Under ideal control, $\mathcal{O}(t)$ exhibits strong dynamical fluctuations and deviates from the target state, while optimized control suppresses these fluctuations and drives the system toward the target state.
	Figure~\ref{Fig.2}(g) shows the adiabatic fidelity as a function of the rescaled time $t/T_{\mathrm{tot}}$. Ideal pulse control gives final-state fidelity $\sim0.97$, while optimization improves it to $\sim0.995$. In the approximately Markovian regime ($\gamma=20$)~\cite{rivas2014quantum}, the method still achieves high fidelity enhancement, demonstrating strong generalization across different environmental regimes. The inset of Fig.~\ref{Fig.2}(g) shows the corresponding control pulses, which exhibit multi-frequency modulation patterns, consistent with the presence of non-adiabatic transitions and dissipative effects~\cite{glaser2015training,brif2010control}. 
	Figure~\ref{Fig.2}(h) shows the MSE averaged over local observables, which stays below  $4\times 10^{-5}$ throughout evolution, demonstrating high prediction accuracy.

	\begin{figure}[htbp]
		\centering
		\includegraphics[width=\linewidth]{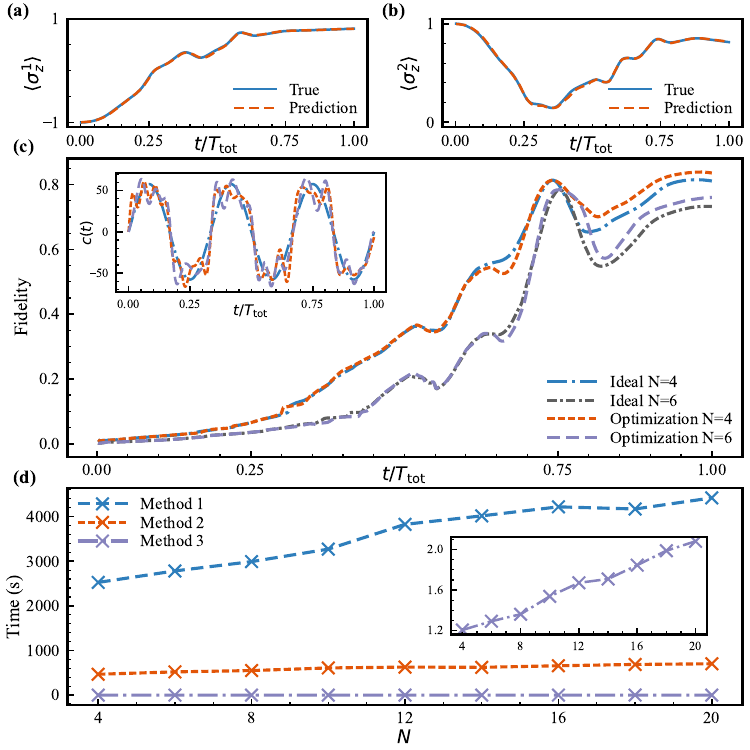}
		\caption{
			Quantum state transfer in an open one-dimensional XY spin chain ($\Gamma=0.3$, $\gamma=10$, $T=10$, $T_{\mathrm{tot}}=\pi/4$). 
			(a)–(b) Local observables $\langle \sigma_z^1 \rangle$ and $\langle \sigma_z^2 \rangle$ for $N=4$.
			(c) The transmission fidelity under ideal and optimized control for $N=4, \gamma=10$ and $N=6, \gamma=15$. The inset shows the corresponding control pulses; 
			(d) Runtime scaling of different methods with system size $N$. Method 1: RK4-based dynamics with iterative control optimization; Method 2: LSTM-based dynamics with iterative control optimization; Method 3: end-to-end quantum control based on latent dynamical manifold learning.}
		\label{Fig.4}
	\end{figure}
	
	We further examine the parameter generalization and stability of the proposed framework~\cite{dong2023learning,lee2024robust} under multi-parameter, time-varying environments, covering both training and out-of-training regimes, with detailed configurations provided in SM~S4--S5. We consider stochastic fluctuations in the environmental parameters, $p(t)\rightarrow p(t)[1+\mathrm{rand}(\lambda)]$, where $p=\Gamma,\gamma,T$ and $\lambda$ denotes the noise amplitude~\cite{wang2024adaptive}. To quantify the improvement achieved by optimized control, we define the  fidelity $F_{\mathrm{im}}=F^{\mathrm{opt}}(T_{\mathrm{tot}})-F^{\mathrm{ideal}}(T_{\mathrm{tot}}).$ Figure~\ref{Fig.3}(a) shows $F_{\mathrm{im}}$ as a function of $\Gamma$, $T$, and $\gamma$. As these parameters increase, $F_{\mathrm{im}}$ exhibits an approximately linear growth, indicating that stronger environmental perturbations provide greater room for control improvement, consistent with Refs.~\cite{Xie2022,gao2026supervised}. 
	For the $\gamma$ scan, RK4 benchmarks and results under time-varying noise ($\lambda=0.05$) remain in close agreement, demonstrating stable performance under time-varying environments.
	Figure~\ref{Fig.3}(b) shows the MSE averaged over environmental parameters under different noise amplitudes. Although the error increases slightly with noise strength, it remains below $4\times10^{-5}$ even at $\lambda=0.1$, confirming that the learned latent dynamics remain stable under environmental fluctuations. Additional MSE results for different parameter configurations are provided in SM~S5. Beyond parameter variations, in SM S6, we further demonstrate generalization across a variety of adiabatic state-preparation tasks in the same open two-level quantum system. By incorporating an Initial-State Encoder~\cite{an2025dual} and treating the driving field $s(t)$ as part of the model input~\cite{zhong2026optimal}, the framework remains effective under varying initial conditions and driving fields. In contrast to PINNs, which encode governing equations as soft constraints and exhibit limited extrapolation capability~\cite{karniadakis2021physics}, the proposed framework learns the underlying  control-conditioned representation directly from local observables, while enabling strong generalization beyond the training conditions.
	
	We next consider quantum state transfer in an open one-dimensional XY spin chain as a larger many-body testbed. The system is described by the Hamiltonian $H_s = \sum_{i=1}^{N-1} J_{i,i+1}\left(\sigma_i^x \sigma_{i+1}^x + \sigma_i^y \sigma_{i+1}^y\right)$, where nearest-neighbor coupling strengths $J_{i,i+1}=\sqrt{i(N-i)}$ support perfect state transfer (PST) \cite{christandl2004perfect,christandl2005perfect} in the absence of the environment at time $T_{\mathrm{tot}}=\pi/4$, and $N$ is the length of the chain. Suppose the first and last spin are immersed in the environment \cite{ren2019quantum}, for the dephasing, the Lindblad operator is taken as $L=\sigma_1^z+\sigma_N^z$. The state transfer task considered here is: $|\mathbf{1}\rangle \rightarrow |\mathbf{N}\rangle$, where $|\mathbf{i}\rangle$ denotes the state with the $i$-th spin excited and all others in the ground state. In this case, the dynamics is restricted in the single-excitation subspace. The transmission fidelity $F(t)=\sqrt{\langle \mathbf{N}|\rho(t)|\mathbf{N}\rangle}$ corresponds to the target-site excitation probability, which is directly determined by the local observable $\langle \sigma_z^i \rangle$. For detail, see SM~S2. With LEO control, the system Hamiltonian is $H(t) = H_s(t) + c(t)|\psi_0(t)\rangle \langle \psi_0(t)|,$ where $|\psi_0(t)\rangle=e^{-iH_{\mathrm{PST}} t} |\mathbf{1}\rangle$, $H_{\mathrm{PST}}$ is the system Hamiltonian with PST couplings. Here we take the ideal control as $c^{\mathrm{ideal}}(t)=I\sin(\pi t /\tau)$ with $I \approx 57.7$, $\tau = \pi/24$, satisfying the first root condition of $J_0(I\tau/\pi)$.

	Figure~\ref{Fig.4} demonstrates the scalability and computational efficiency of the proposed framework for a spin chain under noise, with parameters $\Gamma=0.3$, $\gamma=10$, $T=10$, $T_{\mathrm{tot}}=\pi/4$. Figure~\ref{Fig.4}(a)-(b) show the time evolution of local observables $\langle \sigma_z^1 \rangle$ and $\langle \sigma_z^2 \rangle$ for $N=4$, $\gamma=10$, where predictions agree closely with numerical solutions over the full time range. 
	Figure~\ref{Fig.4}(c) shows that the optimized control clearly improves the fidelity, demonstrating scalability in higher-dimensional systems for $N=4, \gamma=10$ and $N=6, \gamma=15$. The inset of Fig.~\ref{Fig.4}(c)  shows the corresponding control functions, where the optimized control pulses exhibit increasingly complex temporal modulation as the system size increases.
	Figure~\ref{Fig.4}(d) shows the runtime scaling of different methods with system size $N$, on an Intel X7 358H CPU with GPU acceleration disabled to ensure a consistent comparison and an interpretable scaling behavior.
	Method 1 employs RK4-based dynamics with iterative control optimization~\cite{Xie2022} (Adam optimizer with 200 epochs) via BPTT on an $N\times N$ density matrix even in the single-excitation subspace, with runtime increasing from $2500$ to $4300$ seconds as $N$ increases from $4$ to $20$.
	Method 2 uses LSTM-based prediction of local observables~\cite{zhong2026optimal} with the same BPTT iterative optimization (Adam, 200 epochs), yielding $450$ to $700$ seconds. 
	In contrast, Method 3 performs end-to-end mapping in the latent space via a single forward pass without iterative optimization or gradient-based control search, with runtime increasing linearly from $1.2$ to $2$ seconds. 
	This reflects a transition from the conventional “simulate-then-optimize” paradigm (Methods 1–2) to a coupled inference framework where dynamics and control are generated jointly in a learned latent representation, resulting in three orders of magnitude speedup and near-linear scaling.
	
	\textit{Conclusion---}We have proposed an end-to-end quantum control framework based on latent dynamical manifold learning that unifies quantum dynamics and control within a low-dimensional representation. The framework enables dynamical prediction and control optimization directly from initial state and environmental conditions, and has been validated on two examples in open-system environments. It further exhibits strong generalization across multi-parameter, time-varying environments and varying initial states and driving fields, in contrast to the limited extrapolation capability reported for PINN-based approaches.
	Compared with conventional iterative approaches, the proposed framework significantly reduces computational cost and improves scalability while maintaining high-fidelity control, substantially alleviating the computational bottleneck of iterative optimization and highlighting the potential of latent representation learning as a unified paradigm for efficient quantum control in complex open systems.
	
	\textit{Acknowledgments---}This work is supported by the Shandong Provincial Natural Science Foundation (Grant No. ZR2024MA046).
	
	\bibliography{ref.bib}
	\clearpage
	\onecolumngrid
	\section{Supplemental Material: ``End to End Learning of Quantum Control on Latent Dynamical Manifold''}

	\twocolumngrid

	\section{S1: Non-Markovian Quantum Master Equations}
	\label{S1}
	
	Consider the dynamics of an open quantum system coupled to  non-Markovian bosonic baths~\cite{breuer2002theory,Strunz1999}, the total Hamiltonian can be written as $H_{\mathrm{tot}} = H_s + H_b + H_{\mathrm{int}},$ where $H_s$, $H_b$, and $H_{\mathrm{int}}$ are the system, bath, and interaction Hamiltonians, respectively. The bosonic environment consists of a collection of harmonic modes, $H_b = \sum_k \omega_k b_k^\dagger b_k ,$ where $\omega_k$ is the frequency of the $k$-th mode, and $b_k$, $b_k^\dagger$ are the annihilation and creation operators satisfying $[b_k,b_k^\dagger]=1.$ The interaction Hamiltonian is taken as $H_{\mathrm{int}}=\sum_k\left(	g_k^* L^\dagger b_k	+	g_k L b_k^\dagger	\right),$ where $L$ is the Lindblad operator and $g_k$ denotes the coupling strength between the system and the $k$-th bath mode. We set $\hbar=1$ throughout.  Within the quantum state diffusion framework~\cite{Strunz1999,Yu2004,Chenyusui2015}, the system wave function is obtained by projecting the total state onto the Bargmann coherent states of the bath
	\begin{equation}
		\begin{aligned}
			\frac{\partial}{\partial t}
			|\psi(t,z_t^*,w_t^*)\rangle
			=
			&
			\Big[
			-iH_s
			+
			L z_t^*
			+
			L^\dagger w_t^*
			-
			L^\dagger \overline{O}(t,z_t^*,w_t^*)
			\\
			&
			-Q
			L \overline{Q}(t,z_t^*,w_t^*)
			\Big]
			|\psi(t,z_t^*,w_t^*)\rangle, 
		\end{aligned}
		\label{eq:S1}
	\end{equation}
	where $z_t^*$ and $w_t^*$ are complex Gaussian noises. The operators $O$ and $Q$, which characterize the environmental memory effects, are defined as
	\begin{equation}
		\begin{aligned}
			\overline{O}(t,z_t^*,w_t^*)
			=
			\int_0^t ds\,
			\alpha(t,s)\,
			O(t,s,z_t^*,w_t^*),
			\\
			\overline{Q}(t,z_t^*,w_t^*)
			=
			\int_0^t ds\,
			\beta(t,s)\,
			Q(t,s,z_t^*,w_t^*),
			\label{eq:S3}
		\end{aligned}
	\end{equation}
	with functional-derivative ansatz~\cite{Strunz1999,Yu2004}
	\begin{equation}
		\begin{aligned}
			O(t,s,z_t^*,w_t^*)
			|\psi(t,z_t^*,w_t^*)\rangle
			=
			\frac{\delta}{\delta z_s^*}
			|\psi(t,z_t^*,w_t^*)\rangle,
			\\
			Q(t,s,z_t^*,w_t^*)
			|\psi(t,z_t^*,w_t^*)\rangle
			=
			\frac{\delta}{\delta w_s^*}
			|\psi(t,z_t^*,w_t^*)\rangle.
			\label{eq:S5}
		\end{aligned}
	\end{equation}
	
	The operators $O$ and $Q$ encode how the state $|\psi(t,z_t^*,w_t^*)\rangle$ at time $t$ depends functionally on the earlier noise realizations $z_s^*$ and $w_s^*$ (with $ s \leq t $), The bath correlation functions are given by
	\begin{equation}
		\begin{aligned}
			\alpha(t,s)
			=
			\int d\omega\,
			J(\omega)
			(\overline{n}_k+1)
			e^{-i\omega(t-s)},
			\\
			\beta(t,s)
			=
			\int d\omega\,
			J(\omega)
			\overline{n}_k
			e^{i\omega(t-s)}
			\label{eq:S7},
		\end{aligned}
	\end{equation}
	where $\overline{n}_k=\frac{1}{e^{\omega_k/T}-1}$is the thermal occupation number, and the bath spectral density~\cite{Yu2004,Wang2021jpa} is chosen as $	J(\omega)=\frac{\Gamma}{\pi}\frac{\omega}{1+(\omega/\gamma)^2}.\label{eq:S8}$
	Here $\Gamma$, $\gamma$, and $T$ denote the system-bath coupling strength,
	the bath character frequency, and temperature respectively. The quantity $1/\gamma$ characterizes the environmental memory time. In particular, the limit $\gamma \rightarrow 0$ corresponds to strong non-Markovian colored noise with long memory, whereas $\gamma \rightarrow \infty$ recovers the Markovian limit.
	
	The reduced density matrix is defined by the ensemble average
	\begin{equation}
		\rho_s
		=
		M[P_t]
		=
		|\psi(t,z_t^*,w_t^*)\rangle
		\langle\psi(t,z_t^*,w_t^*)|,
		\label{eq:S9}
	\end{equation}
	with $ M[\cdot]=\prod_k \frac{1}{\pi} \int d^2 z \, e^{-|z|^2}(\cdot).$ Taking the time derivative gives the non-Markovian master equation~\cite{Yu2004,Chenyusui2015}
	\begin{equation}
		\begin{aligned}
			\frac{\partial}{\partial t}\rho_s
			=
			&
			-i[H_s,\rho_s]
			+
			[L,M[P_t \overline{O}^\dagger(t,z_t^*,w_t^*)]]
			\\
			&
			-
			[L^\dagger,M[\overline{O}(t,z_t^*,w_t^*)P_t]]
			\\
			&
			+
			[L^\dagger,M[P_t \overline{Q}^\dagger(t,z_t^*,w_t^*)]]
			\\
			&
			-
			[L,M[\overline{Q}(t,z_t^*,w_t^*)P_t]].
		\end{aligned}
	\end{equation}
	
	The operators $O$ and $Q$, which contain the stochastic noises $z_t^*$ and $w_t^*$, are generally obtained approximately using perturbative techniques.
	In the weak-coupling limit, only the zeroth-order contributions are retained. Under this approximation, $M[P_t \overline{O}^\dagger(t,z_t^*,w_t^*)]
	=
	\rho_s \overline{O}(t)$, $ 
	M[P_t \overline{Q}^\dagger(t,z_t^*,w_t^*)]
	=
	\rho_s \overline{Q}(t).
	$ 
	The master equation therefore reduces to the closed non-Markovian form
	\begin{equation}
		\begin{aligned}
			\frac{\partial \rho}{\partial t}
			=
			&
			-i[H_s,\rho]
			+
			[L,\rho\overline{O}^{\dagger}]
			-[L^{\dagger},\overline{O}\rho]
			\\
			&
			+
			[L^{\dagger},\rho\overline{Q}^{\dagger}]
			-
			[L,\overline{Q}\rho].
		\end{aligned}
		\label{equ:1}
	\end{equation}
	The auxiliary operators satisfy the coupled equations~\cite{Wang2021jpa}
	\begin{equation}
		\begin{aligned}
			\frac{\partial\overline{O}}{\partial t}
			=
			&
			\left(
			\frac{\Gamma T\gamma}{2}
			-
			\frac{i\Gamma\gamma^{2}}{2}
			\right)L
			\\
			&
			-
			\gamma\overline{O}
			\left[
			-iH_s
			-
			(L^\dagger\overline{O}
			+
			L\overline{Q}),
			\overline{O}
			\right],
		\end{aligned}
		\label{equ:2}
	\end{equation}
	\begin{equation}
		\begin{aligned}
			\frac{\partial\overline{Q}}{\partial t}
			=
			&
			\frac{\Gamma T\gamma}{2}
			L^{\dagger}
			-
			\gamma\overline{Q}
			\\
			&
			+
			\left[
			-iH_s
			-
			(L^\dagger\overline{O}
			+
			L\overline{Q}),
			\overline{Q}
			\right].
		\end{aligned}
		\label{equ:3}
	\end{equation}
	Equations~\eqref{equ:1}--\eqref{equ:3} form a closed system of coupled nonlinear differential equations, which are directly solved using the Runge–Kutta 4th-order (RK4) method in this work.
	
	\section{S2: Leakage-Elimination Operator Control on Quantum Systems}
	\label{S2}
	
	\textit{Leakage-Elimination Operator---}A leakage elimination operator (LEO) Hamiltonian $H_{\mathrm{LEO}}(t)$~\cite{Wu2002,Wang2020a,Wang2020b, WM} has been proposed to eliminate the leakage of the states from an encoded subspace $|\psi_0(t)\rangle$ to other subspaces  $H(t)=H_{0}(t) + H_{\mathrm{LEO}}(t)$, where $H_0(t)$ is the original Hamiltonian and $H_{\mathrm{LEO}}(t)=c(t)|\psi_0(t)\rangle\langle\psi_0(t)|$, $c(t)$ is the control function that describes a sequence of control pulses.
	
	Now we derive the pulse control conditions by using Feshbach $PQ$ partitioning technique \cite{Wunotes}.
	We consider a complete instantaneous basis $\{|\psi_n(t)\rangle\}$ with
	$\langle \psi_m(t)|\psi_n(t)\rangle=\delta_{mn}$.
	Expanding $|\Psi(t)\rangle=\sum_n a_n(t)|\psi_n(t)\rangle$, the Schr\"odinger equation yields
	\begin{equation}
		i\dot{a}_n=\sum_m
		\left[
		\langle\psi_n|H_0(t)|\psi_m\rangle
		-i\langle\psi_n|\dot{\psi}_m\rangle
		\right]a_m.
	\end{equation}
	To confine the dynamics within the subspace spanned by $|\psi_0(t)\rangle$, we employ the Feshbach $PQ$ partitioning technique~\cite{Wang201803}, which decomposes the state vector into the target component $P(t)$ and the orthogonal complement $Q(t)$. This yields a block representation,
	\begin{eqnarray}
		|\Psi(t)\rangle = \begin{bmatrix} P(t) \\ Q(t) \end{bmatrix}, \quad H_0(t) = \begin{bmatrix} h(t) & R(t) \\ W(t) & D(t) \end{bmatrix},
	\end{eqnarray}
	where $h(t)$ and $D(t)$ act within the $P$ and $Q$ subspaces, respectively, while $R(t)$ and $W(t)$ describe inter-subspace coupling. 
	The LEO term modifies only the $P$-subspace diagonal element, $h(t)\rightarrow h'(t)=h(t)+c(t)$. Introducing the rotating-frame amplitude
	\begin{equation}
		p(t)=\exp\!\left[-i\int_0^t h'(s)ds\right]P(t),
	\end{equation}
	the exact elimination of the $Q$ subspace yields a non-Markovian integro-differential equation
	\begin{equation}
		\dot{p}(t) = \int_{0}^{t} \tilde{g}(t,s) p(s) ds,
	\end{equation}
	where the memory kernel $\tilde{g}(t,s)$ is given by
	\begin{equation}
		\begin{aligned}
			\tilde{g}(t,s) =
			&-R(t)\,\mathcal{T}\exp\left[-i\int_{s}^{t} D(s') ds'\right]\\
			&\times W(s)\exp\left[-i\int_{s}^{t} h'(s') ds'\right],
		\end{aligned}
	\end{equation}
	here $\mathcal{T}$ denotes the time-ordering operator, the dynamics can be written in convolution form
	\begin{equation}
		\dot{p}(t) = -\int_{0}^{t} g(t,s)\,
		e^{-i\int_{s}^{t} h'(s') ds'} p(s) ds.
		\label{Eq:kernel_form}
	\end{equation}
	
	\begin{figure}[htbp]
		\centering
		\includegraphics[width=\linewidth]{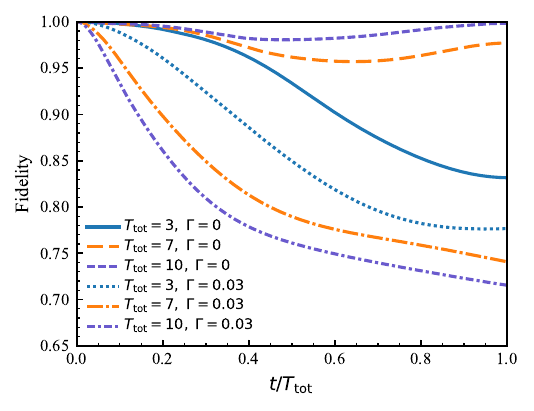}
		\caption{Adiabatic fidelity without control as a function of the rescaled time $t/T_{\mathrm{tot}}$ for different total evolution times $T_{\mathrm{tot}}$, shown for both closed-system ($\Gamma=0$) and open-system ($\Gamma=0.03$) dynamics.}
		\label{SMFig.1}
	\end{figure}
	
	In the strong-driving regime $c(t)\gg h(t)$, the phase factor is dominated by $c(t)$, and we approximate $p(s)\approx p(t)$, since the system evolves slowly compared to the control field.
	Moreover, the memory kernel can be approximated as $g(t,s)\approx g(t-s)$, which typically decays rapidly in $t-s$ and is weakly dependent on microscopic bath details~\cite{PhysRevA.86.032303}. In this regime, replacing $g(t,s)\to g(t,t)$ is also justified. Consequently, within one control period $\tau$, the slowly varying contribution factorizes from the fast phase, yielding the condition
	\begin{eqnarray}
		\label{Eq.(10)}
		\int_{0}^{\tau} ds \exp\left[-i\int_{0}^{s} c(s') ds'\right] = 0.
	\end{eqnarray}
	For rectangular control 
	\begin{eqnarray}
		c(t)=\begin{cases}
			I, & 2n\tau <t<(2n+1)\tau, \\
			-I, & (2n+1)\tau <t<(2n+2)\tau,
		\end{cases} \label{171}
	\end{eqnarray}
	the control condition is
	\begin{eqnarray}
		I\tau =2k\pi, k=1,2,3,... 
	\end{eqnarray}
	
	For sinusoidal control $c(t) = I\sin(\pi t/\tau)$, the condition is determined by the zeros of the Bessel function,
	\begin{equation}
		J_0\!\left(\frac{I\tau}{\pi}\right) = 0,
	\end{equation}
	where $I$ is the pulse amplitude, $\tau$ is the pulse half-period, and $J_0$ is the zeroth-order Bessel function.

	\begin{figure}[htbp]
		\centering
		\includegraphics[width=\linewidth]{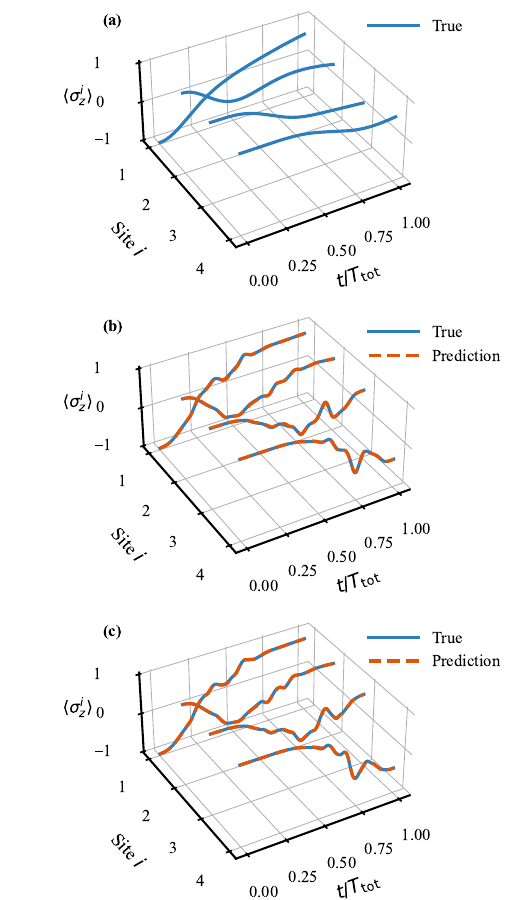}
		\caption{State transfer in a one-dimensional XY spin chain. We show the time evolution of local observables $\langle \sigma_z^i\rangle$ ($i=1,\ldots,4$). Panels (a)–(c) correspond to the cases without control, with ideal control, and with the optimized control generated by the proposed framework, respectively. The system parameters are $N=4$, $\Gamma=0.3$, $\gamma=10$, $T=10$, and $T_{\mathrm{tot}}=\pi/4$.}
		\label{SMFig.2}
	\end{figure}
	
	\textit{Adiabatic speedup in a two-level system---}We consider the adiabatic speedup in an open two-level quantum system, described by the Hamiltonian
	$
	H(t) = [1-s(t)]H_i + s(t)H_f,
	$
	where $s(t)$ is a driving field that linearly interpolates between the initial Hamiltonian $H_i$ and the target Hamiltonian $H_f$. Here we choose $H_i=\sigma_z$, $H_f=\sigma_x$, and the Lindblad operator $L=\sigma_-$, corresponding to dissipative interactions. With LEO control, the system Hamiltonian is modified as
	\begin{equation}
		H(t) = [1+c(t)]\big[(1-s(t))H_i + s(t)H_f\big],
	\end{equation}
	where $c(t)$ is the control field to be optimized. The baseline control is chosen in a sinusoidal $c^{\mathrm{ideal}}(t)=I\sin(\pi t /\tau)$ with amplitude $I \approx 54.4$, half-period $\tau = 1/2$, satisfying the third root condition of $J_0(I\tau/\pi)$ \cite{Wang2020a}. The adiabatic fidelity is defined as $F(t)=\sqrt{\langle E_0(t)|\rho(t)|E_0(t)\rangle}$, where $|E_0(t)\rangle$ is the instantaneous ground state of $H_s(t)$ and $\rho(t)$ is reconstructed from local observables~\cite{nielsen2010quantum}.
	
	In the absence of control, we analyze the adiabatic fidelity for different total evolution time $T_{\mathrm{tot}}$. As shown in Fig.~\ref{SMFig.1}, we consider both the closed-system case ($\Gamma=0$) and the open-system case with ($\Gamma=0.03$). In the absence of dissipation, the adiabatic fidelity increases with increasing $T_{\mathrm{tot}}$, and the system approaches the adiabatic regime for sufficiently large $T_{\mathrm{tot}}$ (e.g., $T_{\mathrm{tot}}\approx 10$). In contrast, in the presence of the environment, the fidelity is progressively suppressed and no longer increases monotonically with $T_{\mathrm{tot}}$. Specifically, for short evolution time ($T_{\mathrm{tot}}=3$), environmental effects are weak and have little impact on the fidelity. For intermediate time ($T_{\mathrm{tot}}=7$), dissipation begins to reduce the fidelity, while for larger time ($T_{\mathrm{tot}}=10$), where the closed system already approaches the adiabatic regime, environmental effects become dominant and significantly degrade the fidelity. This behavior arises from the accumulation of dissipative effects with increasing evolution time.
	
	\textit{State transfer in a one-dimensional XY spin chain---}
	We consider quantum state transfer in an open one-dimensional XY spin chain with Hamiltonian
	\begin{equation}
		H_s = \sum_{i=1}^{N-1} J_{i,i+1}
		\left(\sigma_i^x \sigma_{i+1}^x + \sigma_i^y \sigma_{i+1}^y \right),
	\end{equation}
	where $J_{i,i+1}$ denotes the nearest-neighbor coupling strength between sites $i$ and $i+1$, $N$ is the length of the chain. We choose $J_{i,i+1}=\sqrt{i(N-i)}$, which enables perfect state transfer (PST) in the absence of the environment ~\cite{christandl2004perfect,christandl2005perfect}. We take the model that two end spins of the chain are immersed in the environment \cite{ren2019quantum}, and the Lindblad operator is taken as $L=\sigma_1^z+\sigma_N^z$. In this case, the dynamics can be described within the single-excitation subspace. The simple state transfer task $|\mathbf{1}\rangle \rightarrow |\mathbf{N}\rangle$ is considered here, where $|\mathbf{i}\rangle$ denotes the state with the $i$-th spin excited and all others in the ground state. With leakage elimination control, the Hamiltonian becomes: $H = H_{\mathrm{s}} + H_{\mathrm{LEO}},$ where $H_{\mathrm{LEO}} = c(t)|\psi_0(t)\rangle \langle \psi_0(t)|$ with $|\psi_0(t)\rangle = e^{-iH_{\mathrm{PST}} t} \mathbf{1}\rangle$, $c(t)$ is the control pulse to be optimized~\cite{Xie2022} and $H_{\mathrm{PST}}$ is the Hamiltonian corresponding to PST couplings. Here, the ideal control is given by
	$c^{\mathrm{ideal}}(t)=I\sin(\pi t /\tau)$ with $I \approx 57.7$, $\tau = \pi/24$, satisfying the first root condition of $J_0(I\tau/\pi)$.
	
	\begin{table*}[t]
		\setlength{\tabcolsep}{6pt}
		\centering
		\caption{Network architectures and hyperparameters for different quantum systems.}
		\begin{tabular}{c c c c c c c c}
			\hline
			System & Input dim & Hidden dim & Output dim & LSTM layers & LSTM dropout &  $\mathcal{D}_{\mathrm{dyn}}$ layers &  $\mathcal{D}_{\mathrm{ctrl}}$ layers\\
			\hline
			Two-level system & $10$ & $256$ & $3+K_{\mathrm{ctrl}}$ & $3$ &$0.1$& $1$ & $1$ \\
			1D XY  & $N+5$ & $128$ & $N+K_{\mathrm{ctrl}}$ & $3$ &$0.1$ & $1$ & $1$ \\
			\hline
		\end{tabular}
		\begin{flushleft}
			\footnotesize
			\textit{Note.} \emph{Input dim} and \emph{Output dim} denote the input and output dimensions of the network, respectively. Here, $N$ represents the spin-chain length in the 1D XY model and also corresponds to the number of local observables. \emph{$K_{\mathrm{ctrl}}=M-N+1$} denotes the number of control coefficients generated by the control decoder, corresponding to the control vector $\mathbf{I}$ in Main Text (MT) Eq.~(1), where $k=N,\ldots,M$ specifies the truncation range of the sinusoidal expansion (see Table~\ref{tab:dataset_training} for the values of $N$ and $M$ used in different systems).
			\emph{LSTM depth} denotes the number of stacked recurrent LSTM layers.
			\emph{LSTM dropout} denotes the dropout probability used for regularization of the recurrent layers.
			\emph{$\mathcal{D}_{\mathrm{dyn}}$ layers} and \emph{$\mathcal{D}_{\mathrm{ctrl}}$ layers} denote the number of MLP layers in the dynamical decoder and control decoder, respectively.
		\end{flushleft}
		
		\label{tab:net_config}
	\end{table*}
	
	Under the action of the system Hamiltonian $H_s$ and the chosen Lindblad operators, the dynamics can be restricted to the single-excitation subspace. This is justified by the conservation of the total excitation number $N_{\mathrm{exc}}=\sum_i \sigma_i^+ \sigma_i^-$, which commutes with the XY Hamiltonian $H_s$, ensuring that no transitions occur between different excitation sectors. In addition, the boundary dissipative operator $L=\sigma_1^z+\sigma_N^z$ does not induce excitation creation or annihilation processes. As a result, if the initial state lies in the single-excitation subspace, the entire evolution remains confined to this subspace.
	Within this subspace, the system dynamics can be equivalently represented in the basis $\{|\mathbf{i}\rangle\}$, so that the full $2^N \times 2^N$ density matrix reduces to an effective $N \times N$ representation. To characterize the evolution, we employ local observables $\langle \sigma_z^i \rangle$. The density-matrix diagonal elements directly correspond to site occupation probabilities,
	$
	p_i(t)=\langle \mathbf{i}|\rho(t)|\mathbf{i}\rangle,
	$
	which can be linearly reconstructed from $\langle \sigma_z^i \rangle$ via
	$
	\sigma_z^i = \mathbb{I} - 2\sigma_i^+\sigma_i^-.
	$
	Thus, $\langle \sigma_z^i \rangle$ encodes the excitation occupation at site $i$.
	Accordingly, the state-transfer fidelity is defined as the occupation probability of the target site,
	$
	F(t)=\sqrt{\langle \mathbf{N}|\rho(t)|\mathbf{N}\rangle},
	$
	which corresponds to the final diagonal element of the density matrix in the single-excitation subspace, i.e., the population of the target state $|\mathbf{N}\rangle$, and is equivalently related to $\langle \sigma_z^N \rangle$ via a linear transformation.

	This representation provides a transparent picture of excitation transport in real space. In particular, Figure~\ref{SMFig.2} visualizes the propagation of a single excitation from the first site to the target site at the opposite end of the chain under different control. Here, $\langle \sigma_z^i\rangle=-1$ corresponds to an occupied excitation at site $i$, enabling a direct real-space interpretation of the transport dynamics. In the absence of control, environmental decoherence induces strong spreading of the excitation, which prevents high-fidelity transfer to the target site. Under the ideal control, the excitation is efficiently guided along the chain, resulting in near-perfect state transfer. The optimized control further improves the transfer performance beyond the ideal case, thereby demonstrating the effectiveness of the proposed control strategy in enhancing state-transfer performance under decoherence.
	
	\section{S3: LSTM-based Latent Dynamics with Dual Decoder}
	\label{S3}
	
\begin{figure}[htbp]
		\centering
		\includegraphics[width=\linewidth]{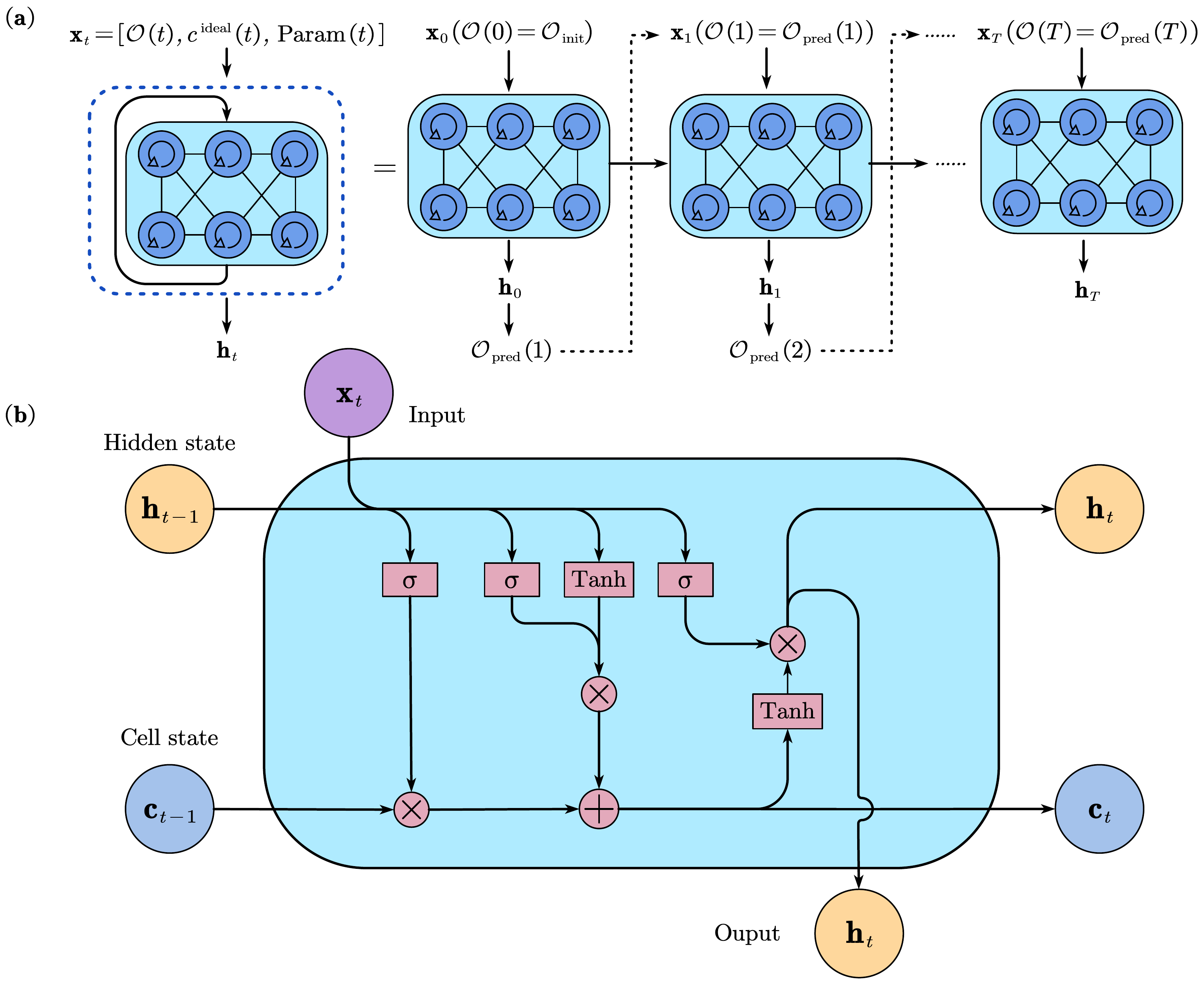}
		\caption{
			Schematic representations of RNNs and LSTM architectures in this framework. 
			(a) Standard RNNs formulation with autoregressive hidden-state evolution, where the observable is generated autoregressively.
			(b) LSTM architecture with gated memory-cell structure enabling stable long-range sequence modeling.}
		\label{SM_LSTM}
	\end{figure}

	\textit{LSTM-based latent representation---}We adopt long short-term memory (LSTM) as the core temporal evolution module in our framework. Figure~\ref{SM_LSTM} illustrated the structure of the LSTM latent dynamical generator. From a structural perspective, LSTM is a specialized recurrent neural networks (RNNs)  for dynamical systems with long-range temporal correlations~\cite{elman1990finding}. 
	As shown in Fig.~\ref{SM_LSTM}(a), the standard RNNs formulation follows an autoregressive hidden-state evolution $(\mathbf{h}_{t+1}, \mathbf{c}_{t+1}) = \mathrm{LSTM}(\mathbf{x}_t, \mathbf{h}_t, \mathbf{c}_t)$,  and the predicted observable $\mathcal{O}_{\mathrm{pred}}(t+1)$ is recursively fed back as part of the next input $\mathbf{x}_{t+1}$. with $\mathcal{O}(0)$ taken as the true initial observable.
	Compared with standard RNNs, LSTM introduces gating mechanisms and an additive cell-state update, enabling stable long-sequence modeling by mitigating vanishing and exploding gradients~\cite{hochreiter1997long}. 
	Its capability originates from its gated memory mechanism, which enables selective suppression of irrelevant historical information and adaptive incorporation of new inputs, thereby achieving a unified representation of multi-scale temporal dependencies~\cite{greff2016lstm}. 
	As illustrated in Fig.~\ref{SM_LSTM}(b), each LSTM unit consists of a hidden state $\mathbf{h}_t$ and a memory cell $\mathbf{c}_t$, governing short-term outputs and long-term dependencies, respectively, with dynamics controlled by input, forget, and output gates~\cite{https://doi.org/10.48550/arxiv.1909.09586, Krichen2025}
	\begin{equation}
		\begin{aligned}
			\begin{bmatrix}
				\mathbf{f}_t \\
				\mathbf{i}_t \\
				\mathbf{o}_t \\
				\widetilde{\mathbf{c}}_t
			\end{bmatrix}
			&=
			\begin{bmatrix}
				\sigma \\
				\sigma \\
				\sigma \\
				\tanh
			\end{bmatrix}
			\!\left(
			W
			\begin{bmatrix}
				\mathbf{x}_t \\
				\mathbf{h}_{t-1}
			\end{bmatrix}
			+
			\mathbf{b}
			\right), \\
			\mathbf{c}_t &= \mathbf{f}_t \odot \mathbf{c}_{t-1}
			+ \mathbf{i}_t \odot \widetilde{\mathbf{c}}_t, \\
			\mathbf{h}_t &= \mathbf{o}_t \odot \tanh\!\left(\mathbf{c}_t\right),
		\end{aligned}
	\end{equation}
	where $\sigma(\cdot)$ denotes the sigmoid activation function, $W$ and $\mathbf{b}$ are the concatenated weight matrix and bias vector corresponding to the forget, input, output, and candidate cell gates, respectively. $\odot$ denotes element-wise multiplication.
	
	\begin{table*}[t]
		\setlength{\tabcolsep}{4pt}
		\centering
		\caption{Dataset construction and training hyperparameters for different quantums systems.}
		\begin{tabular}{c c c c c c c}
			\hline
			System & Samples & Parameter ranges $(\Gamma, \gamma, T)$ & Order $k$ & Epochs & Batch size  \\
			\hline
			Two-level system
			& 12800 
			& $\Gamma \in [0.01,0.05],\gamma \in [1.5,50],T \in [5,15]$ 
			& $k \in [5,30]$ 
			& 100 
			& 128 
			\\
			
			Two-level system(TVN)
			& 12800 
			& $\Gamma \in [0.01,0.05],\gamma \in [1.5,70],T \in [5,15], \lambda \in[0,0.05]$ 
			& $k \in [5,30]$ 
			& 100 
			& 128 
			
			\\
			1D XY($N=4$) 
			& $5120$ 
			& $\Gamma \in [0.25,0.35],\gamma \in [5,15],T \in [10,15]$ 
			& $k \in [15,40]$
			& 120 
			& 64 
			
			\\
			1D XY($N=6$) 
			& $5120$ 
			& $\Gamma \in [0.25,0.35],\gamma \in [5,15],T \in [10,15]$ 
			& $k \in [15,40]$
			& 120 
			& 64 
			\\
			\hline
		\end{tabular}
		\begin{flushleft}
			\footnotesize
			\textit{Note.} 
			\emph{Samples} denotes the number of independent trajectories generated via parameter-space sampling. 
			\emph{Parameter ranges} define the uniform sampling domain used to construct the dataset, where $\Gamma$, $\gamma$, and $T$ denote environmental parameters, and $\lambda$ denotes the noise amplitude in the two-level system under time-varying noise (TVN),
			$p(t)\rightarrow p(t)[1+\mathrm{rand}(\lambda)]$, where $p=\Gamma,\gamma,T$.
			\emph{Order $k$} (integer) specifies the truncation of the control  expansion in MT Eq.1, i.e., the control vector $\mathbf{I}$ consists of coefficients $\mathbf{I}_k$.
			\emph{Epochs} and \emph{batch size}describe the training setup. All models are trained using Adam optimizer with learning rate $lr=10^{-3}$.
		\end{flushleft}
		\label{tab:dataset_training}
	\end{table*}
	
	From a physical perspective, the proposed architecture is particularly suitable for modeling open quantum systems, which often exhibit pronounced non-Markovian dynamics due to finite environmental memory effects. 
	In this sense, the LSTM can be viewed as a latent representation model~\cite{lusch2018deep, rubanova2019latent, krishnan2015deep}, which embeds quantum trajectories into a low-dimensional latent space through recurrent gated dynamics, thereby yielding a structured latent representation $\mathcal{M}$. Within this representation, the hidden-state sequence $\mathbf{h}_t$ forms a trajectory in $\mathcal{M}$, such that different dynamical processes are organized in a smooth and separable manner.
	Furthermore, control information is recursively integrated through the cell-state dynamics and co-evolves with the system representation, yielding a unified representation of dynamics and control. Consequently, the final hidden state provides a global representation of the controlled evolution, while gradients can be propagated through the entire trajectory via backpropagation through time (BPTT)~\cite{werbos1990backpropagation}, enabling end-to-end joint optimization of both the latent dynamics and the control parameters.
	
	\textit{Dual decoder---}Built upon the LSTM-based latent dynamical generator, we introduce a dual-decoder architecture to disentangle dynamical and control-relevant information encoded in $\mathcal{M}$. Specifically, both the dynamical decoder and the control decoder are implemented as multilayer perceptrons (MLP) with identical feed-forward structure~\cite{hornik1989multilayer}, acting on different latent-state interfaces.
	Given an input latent state $\mathbf{u}\in\{\mathbf{h}_t,\mathbf{h}_T\}$, the general form of the decoder is given by
	\begin{equation}
		\mathbf{z}^{(l)}=\phi\!\left(\mathbf{W}^{(l)}\mathbf{z}^{(l-1)}+\mathbf{b}^{(l)}\right),	\mathcal{D}(\mathbf{u})=\mathbf{W}^{(N)}\mathbf{z}^{(N-1)}+\mathbf{b}^{(N)},
		\label{decoder}
	\end{equation}
	here $\mathcal{D}(\cdot)$ denotes either the dynamical decoder or the control decoder, and $N$ is the number of MLP layers. The parameters $\{\mathbf{W}^{(l)}, \mathbf{b}^{(l)}\}_{l=1}^{N}$ denote the layer-wise trainable weight matrices and bias vectors, respectively. The index $l$ labels the hidden layers, with $\mathbf{z}^{(0)}=\mathbf{u}$ as the input latent state and $\mathbf{z}^{(N)}=\mathcal{D}(\mathbf{u})$ as the output. The activation function $\phi(\cdot)$ is applied element-wise to introduce nonlinearity. 
	This construction defines a feed-forward map from $\mathcal{M}$ to either the observable space (dynamical decoding) or the control parameter space (control decoding). Specifically, when $\mathcal{D}(\cdot)=\mathcal{D}_{\mathrm{dyn}}$, the decoder acts on the local trajectory state $\mathbf{h}_t$, whereas for $\mathcal{D}_{\mathrm{ctrl}}$ it operates on the final latent state $\mathbf{h}_T$, which encodes the full evolution history of the system. Consequently, the dual-decoder architecture provides a structured factorization of the latent representation, enabling simultaneous local dynamical prediction and global control generation while preserving a clear separation between the two tasks.
	
	Table.~\ref{tab:net_config} summarizes the network architectures and hyperparameter configurations for different quantum systems in this framework.
	
	\section{S4: End-to-End Training and Gradient Flow}
	\label{S4}
	
	\textit{Staged training strategy---}To ensure a stable latent dynamical manifold while avoiding interference from control optimization on under-converged dynamical representations, we adopt a staged training strategy~\cite{bengio2009curriculum}. Since control synthesis fundamentally relies on the latent representation of system dynamics, the model is first required to learn a reliable and generalizable dynamical manifold before enabling control optimization via control loss $\mathcal{L}_{\mathrm{ctrl}}$.
	
	\begin{figure}[htbp]
		\centering
		\includegraphics[width=\linewidth]{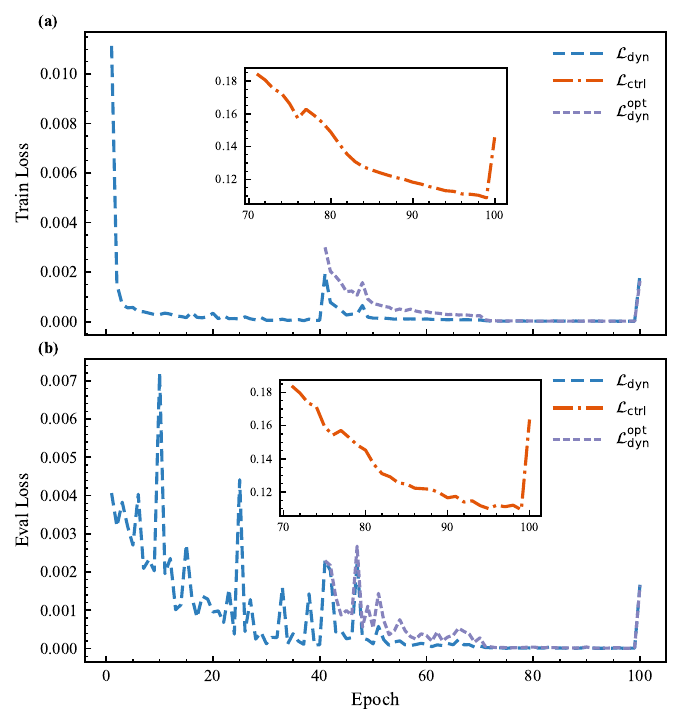}
		\caption{
			Training and evaluation loss evolution for a two-level quantum system under the proposed staged learning strategy. 
			The model is trained sequentially using the dynamical prediction loss $\mathcal{L}_{\mathrm{dyn}}$, the closed-loop dynamical consistency loss $\mathcal{L}_{\mathrm{dyn}}^{\mathrm{opt}}$, and the control optimization loss $\mathcal{L}_{\mathrm{ctrl}}$. 
			Panels (a) and (b) show the training and evaluation losses, respectively. The activation of $\mathcal{L}_{\mathrm{ctrl}}$ at a later stage is indicated in the inset.
		}	
		\label{SMFig.3}
	\end{figure}
	
	In the first stage, only the dynamical prediction loss $\mathcal{L}_{\mathrm{dyn}}$ is optimized. Trained on supervised dynamical trajectories, the model establishes a stable mapping from the observation space to $\mathcal{M}$ under ideal control conditions, while learning the coupling between environmental parameters and control fields with system evolution, thereby forming a latent representation that encodes dynamical evolution and its control dependence.
	
	In the second stage, we introduce the closed-loop dynamical consistency loss $\mathcal{L}_{\mathrm{dyn}}^{\mathrm{opt}}$. Control coefficient vector $\mathbf{I}$, see MT Eq.1 for detail, are randomly sampled from the original control sequence to generate perturbed control fields, and corresponding supervisory trajectories are computed using an RK4 solver. The model is trained to match predictions under perturbed control conditions with numerical integration results. This procedure introduces a control-induced distribution shift~\cite{quinonero2008dataset}, enforcing stability of the latent manifold under control conditions beyond the training distribution, thereby improving reliability for subsequent control optimization.
	
	In the third stage, the control optimization loss $\mathcal{L}_{\mathrm{ctrl}}$ is introduced. The control decoder outputs a control coefficient vector $\mathbf{I}$ from the final latent state, which parameterizes the control pulse $c^{\mathrm{opt}}(t)$ according to MT Eq.1 and then fed back into the differentiable RK4 solver for quantum evolution~\cite{chen2018neural,rackauckas2020universal}, yielding the corresponding fidelity and thus  $\mathcal{L}_{\mathrm{ctrl}}$. 
	Importantly, in this stage all loss terms are jointly optimized, with gradients from $\mathcal{L}_{\mathrm{dyn}}$, $\mathcal{L}_{\mathrm{dyn}}^{\mathrm{opt}}$, and $\mathcal{L}_{\mathrm{ctrl}}$ simultaneously propagating through the shared latent backbone. This control optimization operates on a well-established latent manifold, thereby enabling joint learning of dynamics modeling and control synthesis.
	
	\begin{figure*}[htbp]
		\centering
		\includegraphics[width=\linewidth]{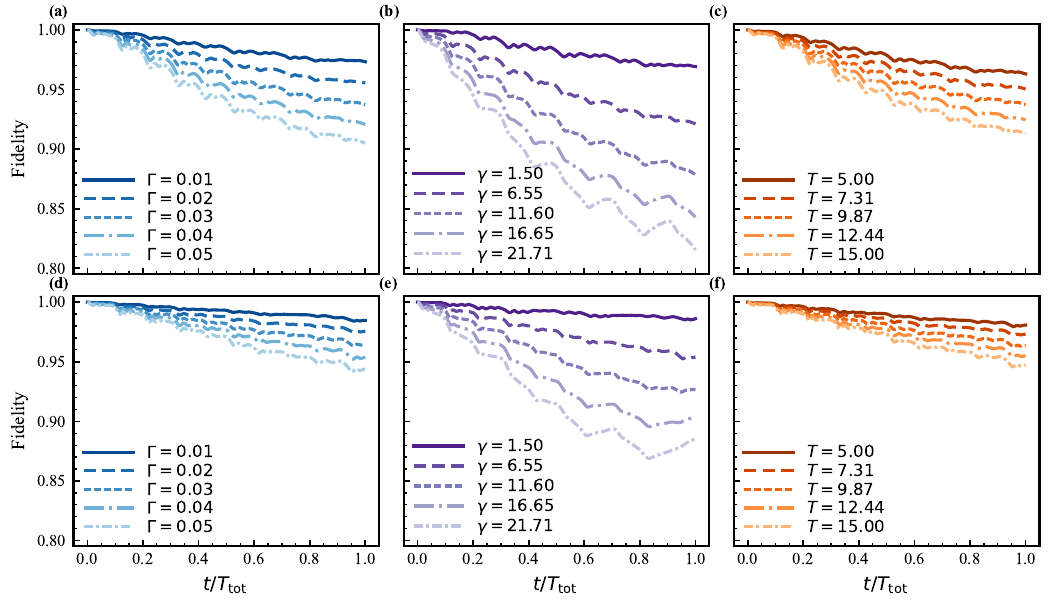}
		\caption{
			Adiabatic fidelity evolution for a two-level open quantum system under time-varying noise ($\lambda=0.05$). Panels (a)–(c) show results under ideal control, while panels (d)–(f) correspond to optimized control. In all cases, single-parameter scans are performed over $\Gamma$, $\gamma$, and $T$, respectively. The remaining parameters are fixed at $\Gamma=0.03$, $\gamma=4$, $T=10$, and $T_{\mathrm{tot}}=5$.
		}
		\label{SMFig.4}
	\end{figure*}
	
	The staged learning procedure is illustrated in Fig.~\ref{SMFig.3}, which shows the evolution of both training and evaluation losses for the two-level quantum system. During the first stage, only $\mathcal{L}_{\mathrm{dyn}}$ is active and decreases steadily. 
	Upon introducing $\mathcal{L}_{\mathrm{dyn}}^{\mathrm{opt}}$ in the second stage, a slight increase in $\mathcal{L}_{\mathrm{dyn}}$ is observed due to the control-induced distribution shift. However it enhances closed-loop dynamical consistency, enabling faster convergence during subsequent joint optimization.
	In the third stage, $\mathcal{L}_{\mathrm{ctrl}}$ is activated and all objectives are jointly optimized, leading to rapid convergence of the control objective. 
	
	The dataset construction and training hyperparameters for all quantum systems considered in this work are summarized in Table.~\ref{tab:dataset_training}.

	\textit{Gradient propagation pathways---}The staged optimization defines a fully differentiable computational graph~\cite{baydin2018automatic}, where gradients propagate across the observation space, latent manifold, and control parameter space in a unified manner, yielding the total loss~\cite{niu2019universal}
	\begin{equation}
		\mathcal{L}_{\mathrm{tot}}=\mathcal{L}_{\mathrm{dyn}}+\mathcal{L}_{\mathrm{ctrl}}+\mathcal{L}^{\mathrm{opt}}_{\mathrm{dyn}}.
	\end{equation}
	
	For the dynamical prediction loss $\mathcal{L}_{\mathrm{dyn}}$ and closed-loop dynamical consistency loss $\mathcal{L}^{\mathrm{opt}}_{\mathrm{dyn}}$, we denote either case generically as $\mathcal{L}_{\mathrm{dyn}}^{dr}$, gradients are first propagated from the local observable mean squared error (MSE) through the dynamical decoder to the latent state
	\begin{equation}
		\frac{\partial \mathcal{L}_{\mathrm{dyn}}^{dr}}{\partial \mathbf{h}_t}
		=
		\frac{\partial \mathcal{L}_{\mathrm{dyn}}^{dr}}{\partial \mathcal{O}_{\mathrm{pred}}(t)}
		\cdot
		\frac{\partial \mathcal{O}_{\mathrm{pred}}(t)}{\partial \mathbf{h}_t},
	\end{equation}
	Here, $\frac{\partial \mathcal{O}_{\mathrm{pred}}}{\partial \mathbf{h}_t}$ is implemented by the dynamical decoder $\mathcal{D}_{\mathrm{dyn}}$. This gradient is further propagated through the temporal recurrent structure of the LSTM, yielding BPTT~\cite{werbos1990backpropagation} along the latent trajectory 
	$
	\frac{\partial \mathcal{L}_{\mathrm{dyn}}^{dr}}{\partial \mathbf{h}_{t-1}}
	=
	\frac{\partial \mathcal{L}_{\mathrm{dyn}}^{dr}}{\partial \mathbf{h}_t}
	\frac{\partial \mathbf{h}_t}{\partial \mathbf{h}_{t-1}}.
	$
	Consequently, gradients accumulated over all time steps propagate into the parameter space, jointly updating the latent dynamical parameters and the decoder parameters
	\begin{equation}
		\frac{\partial \mathcal{L}_{\mathrm{dyn}}^{dr}}{\partial \Theta}
		=
		\sum_{t=1}^{T}
		\frac{\partial \mathcal{L}_{\mathrm{dyn}}^{dr}}{\partial \mathbf{h}_t}
		\frac{\partial \mathbf{h}_t}{\partial \Theta},
		\quad
		\Theta \equiv \{\theta_{\mathrm{latent}}, \theta_{\mathcal{D}_{\mathrm{dyn}}}\}.
	\end{equation}
	The optimization of $\theta_{\mathrm{latent}}$ promotes a dynamics-consistent structure in the latent representation $\mathcal{M}$~\cite{champion2019data}, ensuring that the generated latent trajectories are consistent with the observed quantum dynamics and enabling the model to learn a control-conditioned representation directly from local observables.
	In contrast, $\theta_{\mathcal{D}_{\mathrm{dyn}}}$ governs the mapping from the latent representation to the physical observable space, ensuring that the embedding faithfully reconstructs the quantum dynamical evolution.
	
	For the control loss $\mathcal{L}_{\mathrm{ctrl}}$, gradients are first propagated to the final latent state $\mathbf{h}_T$. This process can be written as
	\begin{equation}
		\frac{\partial \mathcal{L}_{\mathrm{ctrl}}}{\partial \mathbf{h}_T}
		=
		\frac{\partial \mathcal{L}_{\mathrm{ctrl}}}{\partial F}
		\cdot
		\frac{\partial F}{\partial \mathcal{O}(T)}
		\cdot
		\frac{\partial \mathcal{O}(T)}{\partial c^{\mathrm{opt}}(t)}
		\cdot
		\frac{\partial c^{\mathrm{opt}}(t)}{\partial \mathbf{I}}
		\cdot
		\frac{\partial \mathbf{I}}{\partial \mathbf{h}_T},
	\end{equation}
	here $F$ denotes the fidelity between the target quantum state and the final state obtained under the optimized control $c^{\mathrm{opt}}(t)$. The dependence of $\mathcal{O}(T)$ on $c^{\mathrm{opt}}(t)$ is implemented via a differentiable RK4 integration scheme~\cite{chen2018neural,rackauckas2020universal}. The mapping $\frac{\partial \mathbf{I}}{\partial \mathbf{h}_T}$ is realized by $\mathcal{D}_{\mathrm{ctrl}}$, which projects the final latent state onto the control coefficient vector  $\mathbf{I}$.
	Subsequently, the gradient at $\mathbf{h}_T$ is propagated backward through the recurrent dynamics via BPTT, yielding 
	$
	\frac{\partial \mathcal{L}_{\mathrm{ctrl}}}{\partial \mathbf{h}_{t-1}}
	=
	\frac{\partial \mathcal{L}_{\mathrm{ctrl}}}{\partial \mathbf{h}_t}
	\frac{\partial \mathbf{h}_t}{\partial \mathbf{h}_{t-1}}.
	$
	Therefore, the control-induced gradients propagate across the entire temporal sequence and ultimately flow into the parameter space, leading to joint updates of the model parameters
	\begin{equation}
		\frac{\partial \mathcal{L}_{\mathrm{ctrl}}}{\partial \Theta}
		=
		\sum_{t=1}^{T}
		\frac{\partial \mathcal{L}_{\mathrm{ctrl}}}{\partial \mathbf{h}_t}
		\frac{\partial \mathbf{h}_t}{\partial \Theta},
		\quad
		\Theta \equiv \{\theta_{\mathrm{latent}}, \theta_{\mathcal{D}_{\mathrm{ctrl}}}\}.
	\end{equation}
	In this process, $\theta_{\mathrm{latent}}$ governs the response structure of the latent dynamics to control signals, determining how intrinsic dynamical information is jointly represented with its coupling to external control. In contrast, $\theta_{\mathcal{D}_{\mathrm{ctrl}}}$ constrains the mapping from the latent manifold to the  control vector, ensuring the physical consistency and validity of the generated control.
	
	Overall, the optimization procedure admits a unified computational graph interpretation, where the three loss terms define distinct but coupled gradient propagation pathways	
	\begin{flalign*}
		&\mathcal{L}_{\mathrm{dyn}} \rightarrow \mathcal{O}(t) \rightarrow \mathcal{D}_{\mathrm{dyn}} \rightarrow \mathbf{h}_t \rightarrow \theta_{\mathrm{latent}}, \\
		&\mathcal{L}_{\mathrm{ctrl}} \rightarrow F \rightarrow \mathcal{O}(T) \rightarrow c^{\mathrm{opt}}(t) \rightarrow \mathcal{D}_{\mathrm{ctrl}} \rightarrow \mathbf{h}_T \rightarrow \theta_{\mathrm{latent}}, \\
		&\mathcal{L}_{\mathrm{dyn}}^{ctrl} \rightarrow \mathcal{O}(t) \rightarrow \mathcal{D}_{\mathrm{dyn}} \rightarrow \mathbf{h}_t \rightarrow \theta_{\mathrm{latent}}.
	\end{flalign*}

	\section{S5: Generalization and Stability in Time-Varying Environments}
	\label{sec:S5}
	
	In the MT, we demonstrated that the proposed framework significantly enhances final-state fidelity under varying environmental parameters in an adiabatic speedup protocol for an open two-level quantum system. However, the final fidelity alone does not fully capture the accuracy and stability of the controlled dynamics. We therefore further present the full time-resolved fidelity dynamics and analyze the parameter-generalization capability and stability of the model across both in training and out-of-training regimes.
	To emulate realistic open quantum dynamics, we introduce time-varying noise in all the following cases., modeled as
	$
	p(t)\rightarrow p(t)\,[1+\mathrm{rand}(\lambda)],
	$
	where $p=\Gamma,\gamma,T$ and $\lambda=0.05$ denotes the noise strength, capturing stochastic temporal fluctuations of environmental parameters. All single-parameter scans are performed by varying one parameter among $\Gamma$, $\gamma$, and $T$, while fixing the remaining parameters at $\Gamma=0.03$, $\gamma=4$, $T=10$, and $T_{\mathrm{tot}}=5$.

	\begin{figure}[htbp]
		\centering
		\includegraphics[width=\linewidth]{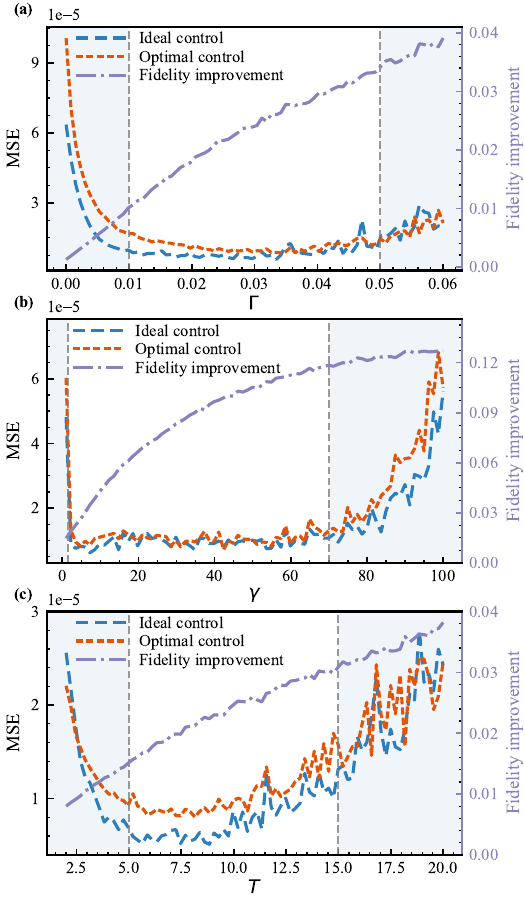}
		\caption{Generalization performance with respect to environmental parameters of a two-level open quantum system under time-varying noise ($\lambda=0.05$), across both in-distribution and out-of-distribution regimes (misty blue shaded). We report the MSE of dynamical predictions under ideal and optimized control, together with the fidelity improvement (right axis, purple). Panels (a)–(c) correspond to single-parameter scans over the environmental parameters $\Gamma$, $\gamma$, and $T$, respectively, while the remaining parameters are fixed at $\Gamma=0.03$, $\gamma=4$, $T=10$, and $T_{\mathrm{tot}}=5$.}
		\label{SMFig.5}
	\end{figure}
	
	Figure~\ref{SMFig.4} illustrates the adiabatic fidelity evolution under variations of $\Gamma$, $\gamma$, and $T$. Panels (a)–(c) show the fidelity under ideal control, while panels (d)–(e) correspond to optimized control trajectories. The results provide a complete characterization of the controlled dynamical process, show that, under both parameter variations and time-varying noise, the system consistently achieves high-fidelity evolution and maintains stable performance over a wide parameter range. This demonstrates the strong generalization and stability of the proposed method against environmental uncertainties.
	
	\begin{figure*}[htbp]
		\centering
		\includegraphics[width=\linewidth]{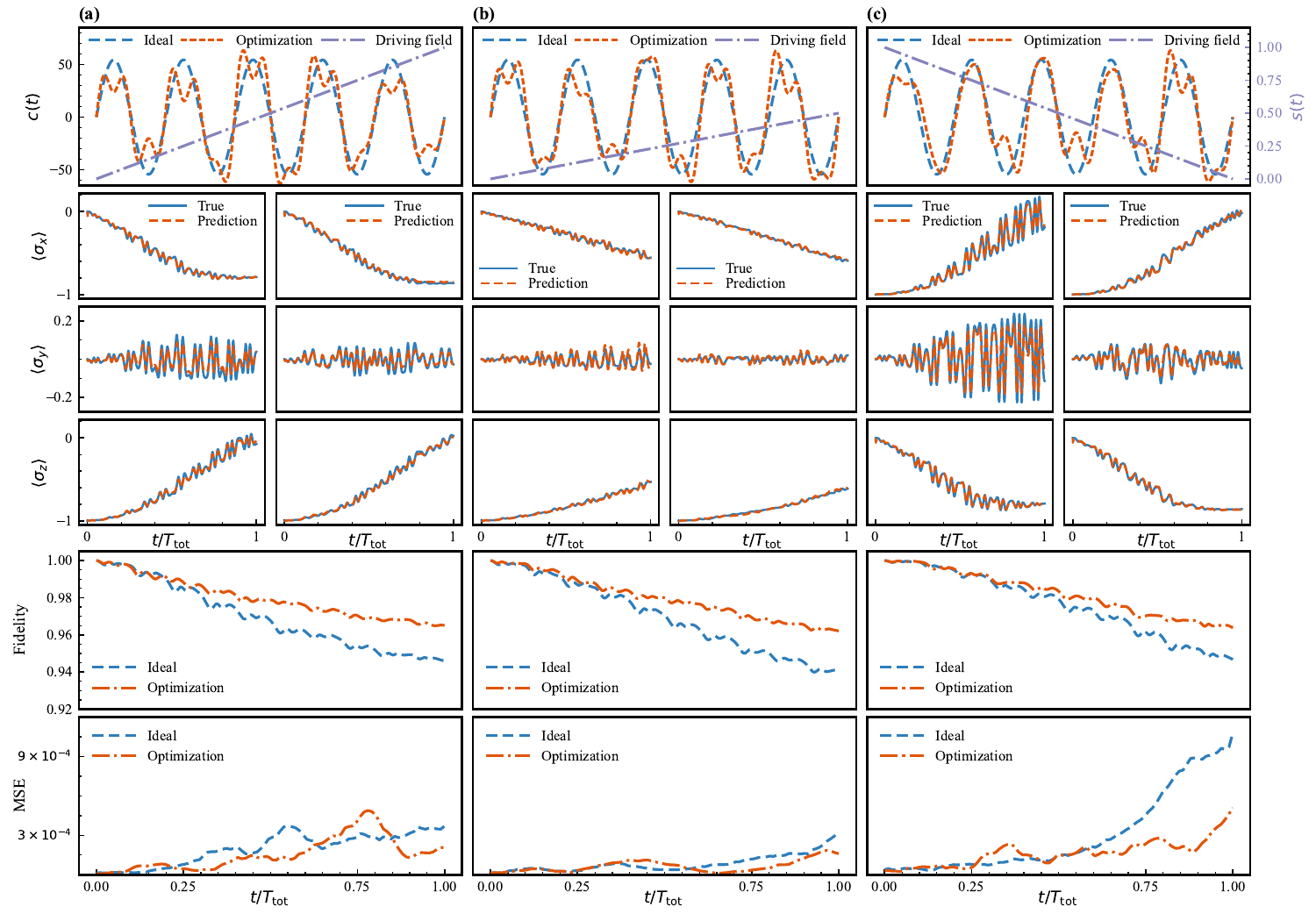}
		\caption{
			Generalization of the proposed framework with respect to initial states and driving field in adiabatic state preparation of a two-level open quantum system.
			(a)–(c) corresponds to a different initial state and driving field $s(t)$.  
			\textit{Top row:} Ideal and optimized control pulses $c(t)$, together with the corresponding driving field $s(t)$ (right axis, purple).
			\textit{Rows 2-4:} Time evolution of local observables $\langle \sigma_{x,y,z} \rangle$ comparing numerical solutions with model predictions. Left: ideal control; right: optimized control.
			\textit{Row 5:}Adiabatic fidelity evolution under ideal and optimized control.   
			\textit{Row 6:} MSE averaged over the observables.
		}
		\label{SMFig.6}
	\end{figure*}
	
	As shown in Fig.~\ref{SMFig.5}, we systematically evaluate the model's generalization behavior under variations of the environmental parameters $\Gamma$, $\gamma$, and $T$. We report the MSE of dynamical predictions under both ideal and optimized control, together with the fidelity improvement induced by control optimization. Panels (a)–(c) correspond to single-parameter scans over $\Gamma$, $\gamma$, and $T$.
	The results show that, under both ideal and optimized control, the model exhibits low dynamical prediction errors near the training distribution center, with increasing errors toward the distribution boundaries, while maintaining comparably low errors in out-of-distribution regimes. The optimized control further yields stable fidelity improvements across all parameter scans. These behaviors persist across both in-distribution and out-of-distribution regimes (misty blue shaded regions), demonstrating strong parameter generalization and stability to environmental uncertainties.
	
	\section{S6. Generalization to Initial States and Driving Field}
	\label{S6}
	
	In this section, we investigate the generalization capability of the proposed framework with respect to different initial-state conditions and driving field $s(t)$~\cite{zhong2026optimal}. 
	For convenience, we construct state-preparation tasks by taking the initial and target states as the ground states of $H(0)$ and $H(T_{\mathrm{tot}})$ determined by the boundary values $s(0)$ and $s(T_{\mathrm{tot}})$, although arbitrary initial and target states can in principle be considered within the framework.
	
	To enable modeling of varying initial-state conditions, we introduce an Initial-State Encoder~\cite{an2025dual}, implemented using the same MLP architecture as the decoders, which embeds the initial quantum state into the initial hidden and cell states of the LSTM:
	$
	(h_0, c_0) = \mathrm{Encoder}(\mathcal{O}_{init}),
	$
	where $\mathcal{O}_{init}$ denotes the initial observable associated with the initial quantum state. 
	We further include the driving field in the input sequence,
	$
	\mathbf{x}_t=[\mathcal{O}(t),\; c(t),\; s(t),\; \mathrm{Param}(t)],
	$
	which enables generalization across different driving fields.
	
	Here, we consider a variety of adiabatic state-preparation tasks in a two-level open quantum system. The system is described by the Hamiltonian
	$
	H(t) = [1 - s(t)] H_i + s(t) H_f,
	$
	where $H_i=\sigma_z$, $H_f=\sigma_x$.
	In the training and evaluation procedures, we consider a variety of linear driving fields
	\begin{equation}
		s(t)=s(0)+\frac{t}{T_{\mathrm{tot}}}\big[s(T_{\mathrm{tot}})-s(0)\big],
	\end{equation}
	which interpolate between different boundary values $s(0)$ and $s(T_{\mathrm{tot}})$. Since these boundary values determine the effective initial and target Hamiltonians, varying them generates different state-preparation tasks. This provides a testbed for assessing the generalization capability of the proposed framework across different initial and target states, as well as different driving fields.
	
	The results in Fig.~\ref{SMFig.6} show that the proposed framework maintains stable dynamical prediction and control optimization across different initial states and driving fields. The model consistently reproduces the trajectories of local observables $\mathcal{O}(t)$ across varying boundary conditions $s(0)$ and $s(T_{\mathrm{tot}})$, demonstrating strong generalization across diverse state-preparation tasks and driving-field configurations. In all cases, $\mathcal{O}(t)$ exhibits strong dynamical fluctuations and deviates from the target state under ideal control, while optimized control suppresses these fluctuations and drives the system toward the target state.
	In particular, the adiabatic fidelity evolution (Row 5) shows that the optimized control consistently achieves higher fidelity than the ideal control across all configurations. Correspondingly, the MSE profiles (Row 6) remain low, confirming accurate dynamical prediction over the entire evolution.
	These results further indicate that the Initial-State Encoder effectively incorporates information from the initial-state into the latent initial condition $(h_0,c_0)$, while the network input and the manifold's generalization constraints enable consistent rollout of dynamics across a broad range of driving fields. 
	Overall, the observed generalization across different initial states and driving fields supports the conclusion that the framework learns a control-conditioned representation from local observables, enabling a unified representation for adiabatic state-preparation control  across different task configurations.
\end{document}